
\input harvmac
 %
\catcode`@=11
\def\rlx{\relax\leavevmode}                  
 %
 %
 %
\font\tenmib=cmmib10
\font\sevenmib=cmmib10 at 7pt 
\font\fivemib=cmmib10 at 5pt  
\font\tenbsy=cmbsy10
\font\sevenbsy=cmbsy10 at 7pt 
\font\fivebsy=cmbsy10 at 5pt  
\def\BMfont{\textfont0\tenbf \scriptfont0\sevenbf
                              \scriptscriptfont0\fivebf
            \textfont1\tenmib \scriptfont1\sevenmib
                               \scriptscriptfont1\fivemib
            \textfont2\tenbsy \scriptfont2\sevenbsy
                               \scriptscriptfont2\fivebsy}
\def\BM#1{\rlx\ifmmode\mathchoice
                      {\hbox{$\BMfont#1$}}
                      {\hbox{$\BMfont#1$}}
                      {\hbox{$\scriptstyle\BMfont#1$}}
                      {\hbox{$\scriptscriptstyle\BMfont#1$}}
                 \else{$\BMfont#1$}\fi}
 %
 %
 %
 %
\def\inbar{\vrule height1.5ex width.4pt depth0pt}
\def\sinbar{\vrule height1ex width.35pt depth0pt}
\def\ssinbar{\vrule height.7ex width.3pt depth0pt}
\font\cmss=cmss10
\font\cmsss=cmss10 at 7pt
\def\ZZ{\rlx\leavevmode
             \ifmmode\mathchoice
                    {\hbox{\cmss Z\kern-.4em Z}}
                    {\hbox{\cmss Z\kern-.4em Z}}
                    {\lower.9pt\hbox{\cmsss Z\kern-.36em Z}}
                    {\lower1.2pt\hbox{\cmsss Z\kern-.36em Z}}
               \else{\cmss Z\kern-.4em Z}\fi}
\def\Ik{\rlx{\rm I\kern-.18em k}}  
\def\IC{\rlx\leavevmode
             \ifmmode\mathchoice
                    {\hbox{\kern.33em\inbar\kern-.3em{\rm C}}}
                    {\hbox{\kern.33em\inbar\kern-.3em{\rm C}}}
                    {\hbox{\kern.28em\sinbar\kern-.25em{\sevenrm C}}}
                    {\hbox{\kern.25em\ssinbar\kern-.22em{\fiverm C}}}
             \else{\hbox{\kern.3em\inbar\kern-.3em{\rm C}}}\fi}
\def\IP{\rlx{\rm I\kern-.18em P}}
\def\IR{\rlx{\rm I\kern-.18em R}}
\def\Ione{\rlx{\rm 1\kern-2.7pt l}}
 %
 %

 %

\def\intem#1{\par\leavevmode%
              \llap{\hbox to\parindent{\hss{#1}\hfill~}}\ignorespaces}
 %


 %
\newskip\humongous \humongous=0pt plus 1000pt minus 1000pt   
\def\caja{\mathsurround=0pt}
\newif\ifdtup
 %

 %

 %
\def\panorama{\global\dtuptrue \openup2\jot \caja
     \everycr{\noalign{\ifdtup \global\dtupfalse
      \vskip-\lineskiplimit \vskip\normallineskiplimit
      \else \penalty\interdisplaylinepenalty \fi}}}
 %
\def\eqalignno#1{\panorama \tabskip=\humongous
     \halign to\displaywidth{\hfil$\displaystyle{##}$
      \tabskip=0pt&$\displaystyle{{}##}$\hfil
       \tabskip=\humongous&\llap{$##$}\tabskip=0pt\crcr#1\crcr}}
 %

 %

 %

 %
 %
 %
 %
\def\,{\hskip1.5pt}           
 %
\let\a=\alpha
\let\b=\beta
\let\c=\chi

\let\j=\psi                                      

\let\l=\lambda

\let\q=\theta                   \let\Q=\Theta
         
\let\s=\sigma                   \let\S=\Sigma

 %
 %
\def\Box{\sqcap\llap{$\sqcup$}}
\def\lapp{\lower.4ex\hbox{\rlap{$\sim$}} \raise.4ex\hbox{$<$}}
\def\gapp{\lower.4ex\hbox{\rlap{$\sim$}} \raise.4ex\hbox{$>$}}
\def\con{\ifmmode\raise.1ex\hbox{\bf*}
          \else\raise.1ex\hbox{\bf*}\fi}
\def\bo{{\raise.15ex\hbox{\large$\Box\kern-.39em$}}}

\def\dual{\relax\leavevmode\lower.9ex\hbox{\titlerms*}}

\let\8=\otimes
 %
 %
 %
 %

\let\2=\underline
\let\ha=\widehat

 %
\def\dt#1{{\buildrel{\smash{\lower1pt\hbox{.}}}\over{#1}}}

\font\eightrm=cmr8
\def\6(#1){\relax\leavevmode\hbox{\eightrm(}#1\hbox{\eightrm)}}
\def\0#1{\relax\ifmmode\mathaccent"7017{#1}     
                \else\accent23#1\relax\fi}      
\def\7#1#2{{\mathop{\null#2}\limits^{#1}}}      
\def\5#1#2{{\mathop{\null#2}\limits_{#1}}}      
 %

 %

 %

 %

 %
\newbox\t@b@x
\def\rightarrowfill{$\m@th \mathord- \mkern-6mu
     \cleaders\hbox{$\mkern-2mu \mathord- \mkern-2mu$}\hfill
      \mkern-6mu \mathord\rightarrow$}
\def\tooo#1{\setbox\t@b@x=\hbox{$\scriptstyle#1$}%
             \mathrel{\mathop{\hbox to\wd\t@b@x{\rightarrowfill}}%
              \limits^{#1}}\,}
\def\leftarrowfill{$\m@th \mathord\leftarrow \mkern-6mu
     \cleaders\hbox{$\mkern-2mu \mathord- \mkern-2mu$}\hfill
      \mkern-6mu \mathord-$}
\def\froo#1{\setbox\t@b@x=\hbox{$\scriptstyle#1$}%
             \mathrel{\mathop{\hbox to\wd\t@b@x{\leftarrowfill}}%
              \limits^{#1}}\,}
 %
\def\frac#1#2{{#1\over#2}}
\def\frc#1#2{\relax\ifmmode{\textstyle{#1\over#2}} 
                    \else$#1\over#2$\fi}           
 %
\def\Claim#1#2#3{\bigskip\begingroup%
                  \xdef #1{\secsym\the\meqno}%
                   \writedef{#1\leftbracket#1}%
                    \global\advance\meqno by1\wrlabeL#1%
                     \noindent{\bf#2}\,#1{}\,:~\sl#3\vskip1mm\endgroup}

\def\QED{\rlx\hfill$\Box$\kern-7pt\raise3pt\hbox{$\surd$}\bigskip}
 %
 %

\def\K#1#2{\relax\def\normalbaselines{\baselineskip12pt\lineskip3pt
                                       \lineskiplimit3pt}
             \left[\matrix{#1}\right.\!\left\|\,\matrix{#2}\right]}
\def\muthstrut{\vphantom1}
\def\mutrix#1{\null\,\vcenter{\normalbaselines\m@th
        \ialign{\hfil$##$\hfil&&~\hfil$##$\hfill\crcr
            \muthstrut\crcr\noalign{\kern-\baselineskip}
            #1\crcr\muthstrut\crcr\noalign{\kern-\baselineskip}}}\,}

 %
\def\YT#1#2{\vcenter{\hbox{\vbox{\baselineskip0pt\parskip=\medskipamount%
             \def\Box{$\sqcap\llap{$\sqcup$}$\kern-1.2pt}%
              \def\Z{\hfil\vskip-5.8pt}\lineskiplimit0pt\lineskip0pt%
               \setbox0=\hbox{#1}\hsize\wd0\parindent=0pt#2}\,}}}
\def\EU{\rlx\ifmmode \c_{{}_E} \else$\c_{{}_E}$\fi}
\def\TM{\rlx\ifmmode {\cal T_M} \else$\cal T_M$\fi}
\def\TW{\rlx\ifmmode {\cal T_W} \else$\cal T_W$\fi}
\def\CM{\rlx\ifmmode {\cal T\rlap{\bf*}\!\!_M}
             \else$\cal T\rlap{\bf*}\!\!_M$\fi}
\def\hm#1#2{\rlx\ifmmode H^{#1}({\cal M},{#2})
                 \else$H^{#1}({\cal M},{#2})$\fi}
\def\CP#1{\rlx\ifmmode\IP^{#1}\else\IP$^{#1}$\fi}
\def\cP#1{\rlx\ifmmode\IC{\rm P}^{#1}\else$\IC{\rm P}^{#1}$\fi}

\def\sll#1{\rlx\rlap{\,\raise1pt\hbox{/}}{#1}}
\def\Sll#1{\rlx\rlap{\,\kern.6pt\raise1pt\hbox{/}}{#1}\kern-.6pt}
%

 %
 %
\def\ie{\hbox{\it i.e.}}        

\def\CY{Calabi-\kern-.2em Yau}
\def\LGO{Landau-Ginzburg orbifold}
\def\3{\ifmmode\ldots\else$\ldots$\fi}
\def\Z{\hfil\break\rlx\hbox{}\quad}
\def\3{\ifmmode\ldots\else$\ldots$\fi}
\def\?{d\kern-.3em\raise.64ex\hbox{-}}           
\def\9{\raise.43ex\hbox{-}\kern-.37em D}         

 %
 %

 %

 %

\def\NP#1{{\it Nucl.\,Phys.\,}{\bf#1\,}}
\def\PL#1{{\it Phys.\,Lett.\,}{\bf#1\,}}

\def\MPL#1{{\it Mod.\,Phys.\,Lett.\,}{\bf#1\,}}

\def\CMP#1{{\it Commun.\,Math.\,Phys.\,}{\bf#1\,}}

 %
 %
 %
\let\ft=\foot
\noblackbox
\def\SaveTimber{\abovedisplayskip=1.5ex plus.3ex minus.5ex
                \belowdisplayskip=1.5ex plus.3ex minus.5ex
                \abovedisplayshortskip=.2ex plus.2ex minus.4ex
                \belowdisplayshortskip=1.5ex plus.2ex minus.4ex
                \baselineskip=12pt plus1pt minus.5pt
 \parskip=\smallskipamount
 \def\ft##1{\unskip\,\begingroup\footskip9pt plus1pt minus1pt\setbox%
             \strutbox=\hbox{\vrule height6pt depth4.5pt width0pt}%
              \global\advance\ftno by1\footnote{$^{\the\ftno)}$}{##1}%
               \endgroup}
 \def\listrefs{\footatend\vfill\immediate\closeout\rfile%
                \writestoppt\baselineskip=10pt%
                 \centerline{{\bf References}}%
                  \bigskip{\frenchspacing\parindent=20pt\escapechar=` %
                   \rightskip=0pt plus4em\spaceskip=.3333em%
                    \input refs.tmp\vfill\eject}\nonfrenchspacing}}
 %
\def\Afour{\ifx\answ\bigans
            \hsize=16.5truecm\vsize=24.7truecm
             \else
              \hsize=24.7truecm\vsize=16.5truecm
               \fi}
\catcode`@=12


\def\khat{\hat{k}}

\def\phat{\hat{p}}

\def\LG{Landau-Ginzburg}
\def\WP{\rlx{W\rm I\kern-.18em P}}

\def\cp#1#2{\hbox{$\IP_{#1}^{#2}$}}
\def\wp{\cp{(k_1,k_2,k_3,k_4,k_5)}{4}}
\def\cM{\cal M}

\def\Mbar{\bar{\cal M}}

\def\Wbar{\bar{\cal W}}
\def\CY{Calabi-Yau}
\def\Hom{\rm Hom}

\def\bM{\bar{M}}
\def\bN{\bar{N}}
\def\bMR{\bM_{\IR}}
\def\bNR{\bN_{\IR}}

 \def\LGO{Landau-Ginzburg orbifold}
 \def\Xb{\relax\leavevmode\hbox{$X$\kern-.6em%
                   \vrule height.4pt width5.7pt depth-1.8ex}\kern1pt}

     %

\baselineskip=16pt
\def\Afour{\hsize=16.5truecm\vsize=24.7truecm}

\Title{\vbox{\baselineskip12pt   \hbox{IASSNS-HEP-94/38}
                                  \hbox{OSU-M-94-2}}}
{\vbox{\centerline{Mirror Symmetry Constructions: A Review}}}

\centerline{\titlerms Per Berglund} \vskip 1mm
\centerline{\it School of Natural Science} \vskip0mm
\centerline{\it Institute for Advanced Study}       \vskip0mm
 \centerline{\it Olden Lane}  \vskip0mm
 \centerline{\it Princeton, NJ 08540}                 \vskip0mm
 \centerline{\rm berglund\,@\,guinness.ias.edu}              
\vskip .2in
\centerline{\titlerms Sheldon Katz}    \vskip1mm
 \centerline{\it Department of Mathematics}                   \vskip0mm
 \centerline{\it Oklahoma State University} \vskip0mm
 \centerline{\it Stillwater, OK~74078} \vskip0mm
 \centerline{\rm katz\,@\,math.okstate.edu}
\vfill

\centerline{ABSTRACT}\vskip5mm
\vbox{
\baselineskip=14pt
\noindent
We review various  constructions of mirror symmetry in terms
of Landau-Ginzburg orbifolds for arbitrary central charge $c$
and \CY\ hypersurfaces and complete intersections in toric varieties.
In particular it is  shown how the
different techniques are related.}

\vfill
\centerline{{To appear in {\it Essays on Mirror Manifolds II}}}

\Date{\vbox{ \line{May 1994 \hfill}}}
\footline{\hss\tenrm--\,\folio\,--\hss}

\lref\rLib{A.~Libgober and J.~Teitelbaum: `` Lines on \CY\ complete
       intersections, mirror symmetry and Picard-Fuchs equations'',
alg-geom/9301001.}

\lref\rreview{{\it Essays on Mirror
       Symmetry}, ed.\ S.-T.~Yau  (Intl.\ Press,
Hong Kong, 1992).}

\lref\rDixon{For a review and references, see L.~Dixon: in {\it
     Superstrings, Unified Theories and Cosmology 1987},
     eds.~G.~Furlan et al.\ (World Scientific, Singapore, 1988)
     \,p.~67--127.}

\lref\rMax{M.~Kreuzer and H.~Skarke: \CMP{150} (1992) 137.}

\lref\rCdGP{P.~Candelas, X.~ de la Ossa, P.~Green and L.~Parkes: \NP{B359}~
(1991)~21.}

\lref\rCdFKM{P.~Candelas, X.~de la Ossa, A.~Font, S.~Katz,
D.R.~Morrison: ``Mirror Symmetry for Two Parameter Models - I'',
\NP{B416} (1994) 481.}

\lref\rCFKM{P.~Candelas, A.~Font, S.~Katz,
D.R.~Morrison: ``Mirror Symmetry for Two Parameter Models - II'',
University of Texas preprint UTTG-25-93, hepth/9403187.}

\lref\rBatdual{V.~Batyrev: ``Dual polyhedra and mirror symmetry for Calabi-Yau
hypersurfaces in toric varieties'', alg-geom/9310003.}

\lref\rBor{L.A.~Borisov: ``Towards the mirror symmetry for Calabi-Yau
complete intersections in Gorenstein toric Fano varieties'', alg-geom/9310001.}

\lref\rBatBor{V.~Batyrev and L.A.~Borisov: ``Dual cones and mirror symmetry
for generalized Calabi-Yau manifolds'', alg-geom/9402002.}

\lref\rBvS{V.~Batyrev and D.~van Straten: ``Generalized Hypergeometric
Functions and Rational Curves on \CY\ Complete Intersections in Toric
Varieties'',  alg-geom/9307010.}

\lref\rAGM{P.S.~Aspinwall, B.R.~Greene and D.R.~Morrison:
       \PL{B303} (1993) 249, \NP{B416} (1994) 414.}

\lref\rmdmm{P.S.~Aspinwall, B.R.~Greene and D.R.~Morrison:
{\it Internat. Math. Res. Notices} {\bf 1993} 319.}

\lref\rFul{W.~Fulton: {\it Introduction to Toric Varieties},
Annals of Math.\ Studies, vol.~131,
Princeton University Press, Princeton, 1993.}

\lref\rCOK{P.~Candelas, X.~de~la~Ossa and S.~Katz: ``Mirror Symmetry for
Calabi-Yau Hypersurfaces in Weighted $\IP^4$ and an Extension of
Landau-Ginzburg Theory'', in preparation.}

\lref\rPM{P.~Berglund and M.~Henningson: ``Landau-Ginzburg Orbifolds,
Mirror Symmetry and the Elliptic Genus'', IASSNS-HEP-93/92,
hep-th/9401029; see also article in this volume.}

\lref\rLVW{W.~Lerche, C.~Vafa and N.~Warner: \NP{B324} (1989) 427.}

\lref\rVW{C.~Vafa and N.P.~Warner, \PL{218B} (1989) 377\semi
          E.Martinec, \PL{B217} (1989) 431.}

\lref\rWitten{E.~Witten, ``On the Landau-Ginzburg Description of $N=2$
minimal models'', IAS preprint IASSNS-HEP-93/10, hep-th/9304026.}

\lref\rFY{P.~Di~Francesco and S.~Yankielowicz,
\NP{B409} (1993) 186.}

\lref\rGPI{B.R.~Greene and M.R.~Plesser: ``Mirror Manifolds: A Brief Review
and Progress Report'', Cornell preprint CLNS-91-1109, hep-th/9110014\semi
B.R.~Greene, M.R.~Plesser and S.-S.~Roan: ``New Constructions of Mirror
Manifolds: Probing Moduli Space far from Fermat Points'',
in {\it Essays on Mirror Symmetry},  ed.\ S.-T.~Yau  (Intl.\ Press,
Hong Kong, 1992).}

\lref\rGP{B.~R.~Greene and M.~R.~Plesser: \NP{B338}(1990).}

\lref\rBH{P.~Berglund and T.~H\"ubsch, in {\it Essays on Mirror Manifolds},
 ed. S.T.~Yau (International Press, Hong Kong 1992); \NP{B393} (1993) 377.}

\lref\rMPR{P.~Candelas, M.~Lynker and R.~Schimmrigk,
      \NP{B341} (1990) 383.}

\lref\rKSS{M.~Kreuzer, R.~Schimmrigk and H.~Skarke, \NP{B372} (1992) 61.}

\lref\rKSII{M.~Kreuzer and H.~Skarke, \NP{B405} (1993) 305.}

\lref\rKS{M.~Kreuzer and H.~Skarke, \CMP{150} (1992) 137\semi
      A.~Klemm and R.~Schimmrigk~, \NP{B411} (1994) 559\semi
          M.~Kreuzer and H.~Skarke, \NP{B388} (1992) 113.}

\lref\rRoan{S.-S.~Roan, {\it Internat. J. Math.} {\bf 2} (1991) 439.}

\lref\rBK{P.~Berglund and S.~Katz, ``Mirror Symmetry for Hypersurfaces in
Weighted Projective Space and Topological Couplings,
hep-th/9311014 and to appear in \NP{B}.}

\lref\rduality{K.~Kikkawa and M.~Yamasaki,  \PL{B149} (1984) 357 \semi
N.~Sakai and I.~Senda, {\it Prog. Theor. Phys.} {\bf 75} (1984) 692.}

\lref\rphases{E. Witten, \NP{B403} (1993) 159.}

\lref\rKW{S.~Kachru and E.~Witten, \NP{B407} (1993) 637.}

\lref\rGQ{D.~Gepner and Z.~Qiu, \NP{B285} (1987) 423.}

\lref\rVafa{C.~Vafa, \MPL{A4}~(1989)~1615.}

\lref\rArnold{V.I.~Arnold, S.M.~Gusein-Zade and A.N.~Varchenko: {\it
     Singularities of Differentiable Maps, Vol.~I}~ (Birkh\"auser,
     Boston, 1985).}

\lref\rKT{A.~Klemm and S.~Theisen, ``Mirror Maps and Instanton Sums for
Complete Intersections in Weighted Projective Space'', Munich preprint
LMU-TPW-93-08, hep-th/9304034.}

\lref\rMax{M.~Kreuzer, ``The Mirror Map for Invertible LG Models'',
CERN preprint CERN-TH-7165-94, hep-th/9402114.}

\lref\rIV{C.~Vafa: \MPL{A4}~(1989)~1169\semi
K.~Intrilligator and C.~Vafa: \NP{B339}~(1990)~95.}

\lref\rLS{M.~Lynker and R.~Schimmrigk: \PL{249B}~(1990)~237.}

\lref\rKTII{A.~Klemm and S.~Theisen, \NP{B389} (1993) 153.}

\lref\rHKTY{S.~Hosono, A.~Klemm, S.~Theisen and S.T.~Yau: ``Mirror Symmetry,
Mirror Map and Applications to Calabi-Yau Hypersurfaces'', Harvard University
preprint HUMTP-93-0801, hep-th/9308122.}

\lref\rHKTYII{S.~Hosono, A.~Klemm, S.~Theisen and S.T.~Yau: ``Mirror Symmetry,
Mirror Map and Applications to Complete Intersection Calabi-Yau Spaces'',
Harvard University
preprint HUMTP-94/02, hep-th/9406055.}

\lref\rDave{D.R.~Morrison: ``Picard-Fuchs Equations and Mirror Maps For
Hypersurfaces'', in {\it Essays on Mirror
       Symmetry}, ed.\ S.-T.~Yau  (Intl.\ Press,
Hong Kong, 1992).}

\lref\rFont{A.~Font: \NP{B391} (1993) 358.}

\lref\rSheldon{S.~Katz: ``Rational Curves on Calabi-Yau Manifolds: Verifying
Predictions of Mirror Symmetry'', alg-geom/9301006 and to appear in
{\it Algebraic Geometry}, E.~Ballico (ed.), Marcel-Dekker.}

\lref\rGMP{B.R.~Greene, D.R~Morrison and M.R.~Plesser:
``Mirror Manifolds in Higher
Dimension'', Cornell University preprint CLNS-93/1253, hep-th/9402119.}

\lref\rBCOV{M.~Bershadsky, S.~Cecotti, H.~Ooguri and C.~Vafa: ``Holomorphic
Anomaly in Topological Field Theories'', with an appendix by S.~Katz,
Harvard University preprint
HUTP-93/A008, hep-th/9302103, and
to appear in \NP{B}.}

\newsec{Introduction}
\noindent
During recent years $N=2$ superconformal field theories have attained
much interest due to their role as   vacua of the heterotic
string theory.
These classical solutions are of phenomenological interest since they
lead to $N=1$ spacetime supersymmetry in the effective field theory.\ft{It
is enough to require $(0,2)$ worldsheet superconformal invariance and in fact
these models may be of even greater importance since the gauge group is
closer to the Standard Model than is the case for the $(2,2)$ vacua. However,
in this article we will restrict ourselves to the left-right symmetric
$(2,2)$ models.}

Apart from the phenomenological reason of wanting to close the gap
between the low-energy theory of the Standard Model as we know it today and
a string theory at the Planck Scale, there is a more fundamental aspect.
Ultimately one would like to know how the vacuum degeneracy
is broken and why the particular vacuum that corresponds to our universe
is chosen. Although we know very little about these non-perturbative effects
the hope is that we will find certain properties of our perturbative treatment
of string theory which will prevail into the non-perturbative regime.
Mirror symmetry~\ft{For a review see~\rreview.} is believed to be one such
feature.
Being in a certain sense a generalization
of duality transformations~\ft{Recall that duality equates string theory on
a circle of radius $R$ with another theory on a circle of radius
$\a'/R$~\rduality.},
in the sigma model language mirror symmetry relates topologically distinct
target spaces, so called mirror pairs, in such a way that the underlying
physics is the same. To be more precise, at the level of the $(2,2)$
superconformal field theory the relative sign of the left $U(1)$-symmetry in
the $N=2$ algebra is ambiguous and the two different choices lead to isomorphic
theories. At the geometrical level the effect of interchanging the $(c,c)$
ring and $(a,c)$ ring is to swap the elements of $H^{2,1}({\cal M})$ and
$H^{1,1}({\cal W})$. We have here made the choice of identifying the $(c,c)$
($(a,c)$)
ring of marginal operators of charge $(1,1)$ ($(-1,1)$)
with the cohomology class of
complex structure deformations and K\"ahler class deformations respectively.
One effect of the mirror operation is to interchange the spectrum, \ie\
the sign  of the Euler number is flipped.

In trying to understand mirror symmetry we immediately encounter the problem of
given a $(2,2)$ model, how does one construct the mirror model?
Although it is natural to expect such a symmetry based on the observation
that there exists two isomorphic conformal field theories related by
a change of sign of the $U(1)$-symmetry, to explicitly construct a
mirror pair is far from trivial. In fact, to this date, only the original
method by Greene and Plesser~\rGP\ is such that we know that pair of theories
indeed are each
other's mirror manifolds\ft{For a review of the construction of~\rGP\
see the paper by Greene and Plesser in this volume.}. There the models
are made up of tensor products of $N=2$ minimal models, among which are
$c=9$ theories which have been shown to be consistent string vacua and some
of them have, in a certain sense, a geometrical interpretation;
roughly speaking they are the ``small radius'' limit of \CY\ manifolds.

However, there do exist conjectured constructions which go beyond the class
of models considered in~\rGP; conjectured because there does not yet
exist a proof  at the level of the conformal field that they give true mirror
pairs. These are the techniques which will be reviewed
in this article.
Although the constructions to be discussed apply to what one naively may
think of as rather different realizations of $N=2$ models, when put in the
context of the $N=2$ phases picture of Witten~\rphases\ or that of the
enlarged K\"ahler  moduli space, by Aspinwall, Greene and Morrison~\rAGM,
they do in fact apply for the same model but in different regions of the
moduli space.\ft{By enlarged we mean the K\"ahler moduli space with
all its different phases, including the \CY\ phase, the Landau-Ginzburg
orbifold phase and others~\refs{\rphases,\rAGM}.}

The transposition scheme~\rBH,
grew out of the attempts of
understanding mirror symmetry away from the specially symmetric Fermat
points~\refs{\rLS,\rGPI}. It is most naturally thought of in the language
of $N=2$ Landau-Ginzburg orbifolds~\rVW, using the concepts of quantum
and geometric symmetries at the Gepner point, to which it is assumed that
the \LGO\ will flow in the IR-limit. In most of our discussion we will
restrict ourselves to models relevant from a string compactification point
of view, \ie\ one in which the internal theory is made to consist of a
$c=9$ $(2,2)$ superconformal field theory. However, using the transposition
technique mirror pairs can be  constructed for arbitrary central charge~\rPM.

However, there does not necessarily exist a \LGO\ phase in every enlarged
moduli space. Instead we consider mirror families of hypersurfaces in toric
varieties which are studied by means of the Newton polyhedron (and its
polar~\rBatdual),
\ft{For a rather large class of the
geometric constructions  transposition can be straightforwardly
carried out in the large radius limit counterpart of the \LGO\ phase
to  which it was originally applied.}
In addition, a related construction for complete
intersections in toric varieties~\refs{\rBor,\rBatBor} is reviewed.
Rather than associating a particular point in the moduli space of a model
with its counterpart in the moduli space of the mirror, the toric method
naturally gives all of the moduli space, or at least the part which can be
analyzed by toric methods. In comparing the two techniques we will however
restrict ourselves to those parts of the moduli space in which
transposition is valid.

The outline of the paper is as follows. In section~2 we review the
transposition method while the toric method is described in section~3
including a number of examples illuminating the correspondence between
the two approaches. Finally, we discuss some open questions in section~4.

\newsec{$N=2$ Landau-Ginzburg models}
\subsec{General framework}\noindent
Although mirror symmetry has only been proven for the $N=2$ minimal models
(and tensor products or quotients thereof), in terms of trying to generalize
the results of~\rGP\ we are much better off by stating the problem
in terms of the $N=2$ Landau-Ginzburg formalism. Consider the usual
action
\eqn\eACTION{\int d^2z d^2\q d^2\bar\q K(x_i,\bar x_i) +
             (\int d^2z d^2\q W(x_i) + c.c).}
where $K$ is the K\"ahler potential and $W$, the superpotential, is a
holomorphic function of the $N=2$ chiral superfields
$x_i(z,\bar z,\q^+,\q^-)$. Due to
nonrenormalization theorems  $W$ is not renormalized (up to scaling)
and hence will
characterize the theory (modulo irrelevant perturbations coming from
the K\"ahler potential \rVW). Let $W$ be a polynomial in the  superfields
$x_i~,~i=1,\ldots,n$. We require that $W$ is quasi-homogeneous of degree $d$,
 \ie\  under rescaling of the world-sheet
\eqn\eSS{ x_i \mapsto \l^{k_i} x_i~~,\qquad W(x_i) \mapsto \l^d W(x_i)~.}
 To compute the central charge is straightforward \rVW\ ,
 $c=6\sum_{i=1}^n
({1\over 2} - q_i)$ with $q_i=k_i/d$ the charge under the $U(1)~$-
current, $J_0$.
In the case of $W=x^{k+2}$, $c=3k/(k+2)$ and so unless the theory has
a trivial fixed point, it must be  equivalent, in the IR limit, to the $N=2$
minimal model with diagonal invariant at level $k$.\ft{This identification,
originally suggested in~\refs{\rVW,\rLVW} has recently been further
strengthened by comparison of the elliptic genus for the two theories
respectively~\refs{\rWitten,\rFY}.} Of course, the conjectured equivalence
extends to tensor products of {\it all} minimal models,
including the non-diagonal and the exceptional series,
 as well as quotients thereof.

In the case that $\hat c=c/3\in \ZZ$ and the number of superfields is
$\hat c+2$, there is a natural way to associate a sigma model interpretation
where the target space is a \CY\ manifold; the Landau-Ginzburg
orbifold and the \CY\ manifold correspond to two different ``phases''
in the language of~\refs{\rphases,\rAGM}. Roughly speaking, the \CY\ phase
is at large radius, where the manifold picture is a good approximation
of the conformal field theory while the Landau-Ginzburg phase is at small
radius. In terms of constructing honest mirror pairs as equivalent
conformal field theories it is of course important to distinguish
between the different parts of the moduli space. However, at our
current level of understanding of the techniques to be presented below,
we only use the knowledge of the superpotential (or the defining
polynomial) and the quantum/geometric symmetries.\ft{The geometric symmetries
consist of all the symmetries of the superpotential modulo the permutation
symmetries which are considered separately, while the quantum symmetries
act non-trivially on the twisted sectors of the \LGO.} Thus,
in most of the following
discussion we will not distinguish between
the geometric and the Landau-Ginzburg approach.

A natural question to ask is whether mirror symmetry generalizes away from
the $N=2$ minimal models. Based on the original observations~\refs{\rDixon,
\rLVW}\ the answer seems to be
 yes. This is further supported by a classification
of $N=2$ \LGO s (or \CY\ hypersurfaces in weighted projective
space)~\refs{\rMPR,\rKSS,\rKS,\rKSII}
where it was found that the spectrum is to a very high degree
symmetric under interchange of $b_{1,1}$ and $b_{2,1}$, the number of
$(-1,1)$ and $(1,1)$ states
(or $(1,1)$ and $(2,1)$ forms) respectively.
Further evidence is given by comparison between predictions of Yukawa
couplings in the K\"ahler moduli space computed using mirror symmetry and
known geometric results on the intersection numbers and the number of
rational curves~\refs{\rCdGP,\rDave,\rLib,\rFont,\rSheldon,\rKT,\rBvS,\rKTII,
\rCdFKM,\rHKTY,\rBCOV,\rBK,\rGMP,\rCFKM, \rHKTYII}.
Two questions immediately arise:
$1)$ how do we construct a (potential) mirror pair\ft{For $({\cal M,W})$
to qualify as a potential mirror pair, we demand not only that
$b_{1,1}$ and $b_{2,1}$
are interchanged but also that the quantum and geometric symmetries
  get swapped between ${\cal M}$ and ${\cal W}$.
We will not assume
that the two conformal field theories are the same.},
 and $2)$ how do we prove, at the
level of conformal field theory, that the two models indeed correspond to the
same physical theory.
In what follows we will mostly be concerned with $1)$ and will have very
little to say about the second point.

Let us return to the $N=2$ minimal models and see what general features
we can extract, something which will hold even away from these special
models.

First we have to require that the spectrum is interchanged, a necessary
but not sufficient condition. Following~\rIV\ it is straightforward to compute
the number of $(\pm 1,1)$ states for any \LGO. Although slightly more
tedious a similar computation can be done for the corresponding \CY\ when
there exists such a phase in the moduli space. Although not a topological
invariant, the number of massless matter $E_6$ gauge singlets can
also in principle be
calculated in the \LGO\ phase~\rKW. This provides a further check of
the proposed constructions. However, in what follows we will not make
use of this information, in particular because to date this number is
known only for the minimal models for which  there already exists
a proven mirror symmetry construction.

Second, recall that  one of the crucial points in the construction of
mirror pairs for (tensor products) of minimal models is that a quotient by
the full geometric symmetry, $\ZZ_{k+2}$ for an $A_k$ model, gives back the
original theory\ft{This follows from the fact that the minimal models
can be expressed in terms of a $\ZZ_k$ parafermion and a
free boson, where the parafermionic theory is equivalent to a $\ZZ_k$
quotient of itself~\rGQ.}. The resulting theory does not have any
geometric symmetry
left while its quantum symmetry~\rVafa\ is $\ZZ_{k+2}$, \ie\
for a mirror pair ${\cal M},{\cal W}$ we require that
\eqna\eQS
$$\eqalignno{ Q_{\cal M}~~ &\cong ~~G_{\cal W}~, & \eQS a \cr
              G_{\cal M}~~ &\cong ~~Q_{\cal W}~. & \eQS b \cr}$$
where $G_{\cal M}$ ($G_{\cal W}$) and $Q_{\cal M}$ ($Q_{\cal W}$) are
the geometric and quantum symmetries of ${\cal M}$ (${\cal W}$)
respectively. Thus, between the original
model and its maximal quotient we observe that the quantum and geometric
symmetries have been interchanged. Although we  will not be
able to compute the complete partition function and in that way
determine whether the two models form a mirror pair, the above symmetry
argument
is an important guideline when trying to construct mirror pairs. In fact
by studying the $(c,c)$ ($(a,c)$) rings of the two models respectively
it is rather straightforward to check that the $(c,c)$ ring structure for
${\cal M}$ is the same as the $(a,c)$ one for the mirror model ${\cal W}$
and vice versa (up to an overall normalization constant).

\subsec{Fractional transformations}\noindent
A first step in the direction of extending the original mirror construction
is to consider marginal deformations away from the minimal models. In terms
of the Landau-Ginzburg orbifold
representation this implies studying deformations
away from Fermat superpotentials~\rGPI. Although the scaling (quantum)
symmetry has
not changed since the $U(1)$ charges stay fixed the geometric symmetry
does change with the choice of superpotential. The idea is then to construct
the mirror model along the lines described above, \ie\ require that~\eQS{}
are fulfilled and that the spectrum is flipped. To be more precise one
first deforms the theory to the original symmetric, Fermat point~${\cal M}'$.
Now,
we know~\rGP\ that the mirror of ${\cal M}'$, ${\cal W}'$
 is obtained as an appropriate
quotient of ${\cal M}'$. However, because of the different geometric
symmetries that ${\cal M}$ and ${\cal M}'$ are equipped with ${\cal W}'$
cannot be the mirror of ${\cal M}$. But by explicitly performing the
quotient as a fractional transformation~\rLS\ one can indeed obtain the
correct mirror ${\cal W}$.

Following~\rGPI\  let us illustrate these ideas by considering the
 model~\rCdFKM\ ${\cal M}'=p'/j$ where,
\eqn\efermat{
p'~=~x_1^8+x_2^8+x_3^4+x_4^4+x_5^4~.}
and $j=(\ZZ_8:1,1,2,2,2)$~\ft{We use the notation
$(\ZZ_r:\Q_1,\Q_2,\Q_3,\Q_4,\Q_5)$ for a $\ZZ_r$ symmetry
with the action $(y_1,y_2,y_3,y_4,y_5) \to (\a^{\Q_1}
y_1,\ldots,\a^{\Q_5} y_5)$, where $\a^r~=~1$.}
is the usual generalized GSO-projection. Its mirror ${\cal W}'$
is obtained in the usual fashion by constructing the quotient of $p'$
with respect to $(\ZZ_4)^3$. Rather than doing  that we perform a
fractional transformation,
\eqn\eft{
(x_1,x_2,x_3,x_4,x_5)\to (y_1^{3/4},y_2,y_3 y_4^{1/4}, y_4^{3/4} y_5^{1/4},
y_5^{3/4} y_1^{1/4})~,}
The first point is that~\eft\ is not one-to-one unless we perform a $(\ZZ_4)^3$
identification on the $x_i$ and at the same time a $\ZZ_{216}$ identification
on the $y_i$. Secondly, the new model constructed from $p'(x_i)$ after
the change of coordinates, ${\cal W}''=p''/j''$ with
\eqn\eftfer{
p''~=~y_1^6 + y_2^8 + y_3^4 y_4 + y_4^3 y_5 + y_5^3 y_1~,}
and $j''=(\ZZ_{216}:36,27,41,52,60)$
is {\it not} the mirror of $p'$; the quantum and geometric symmetries are
not interchanged. Rather there exists a deformation of $p'$ for which
\eQS{} are fulfilled,
\eqn\edeffer{
p~=~x_1^6 x_5 + x_2^8 + x_3^4 + x_4^3 x_3 + x_5^3 x_4~.}
The construction can be repeated for a number of other special symmetric
points in the complex structure moduli space, and of course for all models
which have a Fermat point~\rGPI.

We now make the following crucial observation; \edeffer\ and \eftfer\ look
very much the same. This is due to the constraint coming from the
quantum/geometric symmetries. In fact, the only difference is the way we have
coupled the terms together, \ie\ which $x_i$ appears with which $x_j$.
Since the models as written have very little to do with the Fermat ones
may guess that in fact there may be a way of constructing mirror pairs
for more general theories. It is to that we turn next.

\subsec{Transposition}\noindent
Although very simple in their structure, as one may suspect most
\LG\ configurations do not  admit a Fermat superpotential~\refs{\rMPR,\rKS}.
 By this
we mean the following. For a given set of fields $x_i$ with
$U(1)$-charges $q_i=k_i/d$ where $k_i$ are the weights and $d$ is the
degree of the superpotential
 there does not exist a point in the  moduli space associated to
deformations by $(1,1)$ operators, \ie\ deformations of the superpotential,
such that
the theory can be represented by a Fermat type superpotential.

The next simplest kind of potentials are of the following
form~\refs{\rArnold,\rMPR,\rKS},
\eqna\etadloop
$$\eqalignno{ p_T&=x_1^{a_1}+x_1x_2^{a_2}+\ldots + x_{n-1}x_n^{a_n}~,
& \etadloop a \cr
              p_L&=x_1^{a_1} x_2 + x_2^{a_2}x_3 + \ldots + x_n^{a_n}x_1~,
& \etadloop b \cr
}$$
denoted tadpole and loop, respectively. The phase symmetry groups
of the two potentials are $G_T\simeq \ZZ_{a_1\cdots a_n}$ and
$G_L\simeq\ZZ_{a_1\cdots a_n+(-1)^{n-1}}$
Note  in particular that both $p_T$ and $p_L$ reduce to the Fermat when $n=1$.
More complicated models can be constructed out of  the tadpole and loop
building blocks by adding two or more potentials together.

Given a model ${\cal M}=p/H$, where we for simplicity take $p=p_T$ and
where $H\subseteq G=G_T$, we want to find its mirror ${\cal W}$. The idea
is to construct another model ${\cal W}$ such that the roles
of the quantum and geometric symmetries are interchanged.
First we
associate to $p$ the matrix of exponents~\rBH\
\eqn\epmat{ A~ = ~\left[
                        \mutrix{ a_1 &  1  &  0  &  \ldots  &  0  \cr
                                  0  & a_2 &  1  &  \ldots  &  0  \cr
                                  0  &  0  & a_3 &  \ldots  &  0  \cr
                             \vdots  &  \vdots  &  \vdots  & \ddots &  1  \cr
                                  0  &  0  &  0  &    & a_n \cr}
                       \right]~, }
whose columns are the degree vectors of the respective monomials of $p$.
The new polynomial is then defined such that its corresponding matrix
is the transpose of the above one, \ie\
\eqn\ephat{
    \ha{p}~~ = ~~y_n^{a_n}  + \ldots
       + y_2^{a_2} y_3 + y_1^{a_1} y_2~~,\quad
    A^{\rm T}~ = ~\left[
                        \mutrix{ a_1 &  0  &  0  &  \ldots  &  0  \cr
                                  1  & a_2 &  0  &  \ldots  &  0  \cr
                                  0  &  1  & a_3 &  \ldots  &  0  \cr
                          \vdots  &  \vdots  &  \vdots  & \ddots &  0  \cr
                                  0  &  0  &  0  &  1  & a_n \cr}
                       \right]~.}
Alternatively, $\ha{p}$ is obtained by reversing the order of the exponents
of $p$. Note that $\ha{p}$, denoted the conjugate or transpose
potential (see~\refs{\rBH,\rPM}),
is of the same type as $p$ and in particular the total phase  symmetry
is the same\ft{Note that $p''$ and $p$ in~\eftfer\ and \edeffer\ are related
by transposition. However, in general $p$ and $\hat p$ do not have to
belong to a Landau-Ginzburg configuration which admits a Fermat potential.}.
The second step is then to find an $\tilde H$ such
that \eQS{}
are fulfilled with ${\cal W}=\ha{p}/\tilde H$. In~\rBH\ it was conjectured
that this is always true, something which was recently proven to be the
case~\rPM.

As an example let us consider the following model ${\cal M}=p/j$, which we
will return to throughout the paper, where
\eqn\epex{
p=x_1^{10} + x_2^5 + x_3^5 + x_4^5 + x_1 x_5^3
}
and $j=(\ZZ_{10}:1,2,2,2,3)$. This model has $(b_{1,1}=3,\,b_{2,1}=75)$ and
$(Q_{\cal M}=\ZZ_{10},\,G_{\cal M}=\ZZ_3\times(\ZZ_5)^3)$.
{}From~\ephat\ we find that the transposed
polynomial is
\eqn\eptrex{
p^T=y_1^{10} y_5 + y_2^5 + y_3^5 + y_4^5 + y_5^3~.
}
The scaling symmetry of $p^T$ is given by $j^T=(\ZZ_{15}:1,3,3,3,5)$
and hence in order to fulfill~\eQS{} the mirror model is given by
${\cal W}=p^T/(j^T\times H)$ where
$H=(\ZZ_5:4,1,0,0,0)\times (\ZZ_5:4,0,1,0,0)$; the number of $(\pm 1,1)$
states are computed in a straightforward manner~\rIV\ and found to be flipped
compared with ${\cal M}$.

It is clear that~\etadloop{} and combinations thereof do not exhaust the list
of possible non-degenerate \LG\ potentials~\rMPR. However, they are the only
polynomials for which $n$ terms (for a theory with $n$ fields) is enough to
ensure that the origin is the only degenerate point. It is obvious that
for the transposition scheme to apply one has to have an $n\times n$ matrix
of exponents. Thus, for the non-invertible models, \ie\ those which are not
made up of the building blocks in~\etadloop{} one will have to discard certain
terms in the superpotential and so the resulting theory is no longer
well-defined. However, we may still go ahead and construct the mirror
as prescribed above. The question is whether the `mirror'  theory makes sense
and if it actually is the mirror.
In the following section we will argue that this indeed is the case.

\newsec{The toric approach}
\subsec{Toric Generalities}
\noindent
For the convenience of the reader we quickly summarize the key facts about
toric varieties which we will make use of.  For more details, see~\rFul.

A toric variety $X$ is a complex variety admitting an action of
the algebraic torus $T=(\IC^*)^r$, $X$ containing a dense open subset
isomorphic to $T$
such that the translation action of $T$ on itself extends holomorphically
to an action of $T$ on $X$.

There are two lattices that arise frequently in the theory of toric
varieties:  the lattice $N=\Hom(\IC^*,T)$ of one~parameter subgroups of $T$,
and the dual lattice $M=\Hom(T,\IC^*)$ of characters of $T$.  Conversely,
starting with a lattice $M$ abstractly isomorphic to $\ZZ^r$, we can
intrinsically describe $T$ as $\Hom(M,\IC^*)$ (or in terms of $N$,
we get $T=N\otimes\IC^*$).  We put $N_{\IR}=N\otimes\IR$ and
$M_{\IR}=M\otimes\IR$.

Toric varieties are commonly described in two ways: by a fan in $N_{\IR}$, or
by a polyhedron in $M_{\IR}$.

A {\it fan\/} $\S$ is a finite collection of rational strongly
convex polyhedral cones
$\s\subset N_{\IR}$, such
that if $\s\in\S$, then each face of $\s$ is also in
$\S$, and finally that if $\s,\s'$ are in $\S$, then $\s\cap\s'$ is a face
of each.  In other words, each cone is spanned by finitely
many edges of the form $\IR_{\ge 0}\cdot n_i$ with $n_i\in N_{\IR}$
rational relative
to the lattice $N$; the adjective ``strongly''
implies that $\s$ contains no nonzero linear subspace of
$N_{\IR}$.
Each cone $\s\in\S$ determines an affine toric variety $U_{\s}$, and
the $U_{\s}$ glue along their intersections $U_{\s\cap\s'}$ to form the
toric variety $X_{\S}$.  Let $\S(1)$ denote the set of
the edges of $\S$.  We will often abuse notation by denoting an edge by its
{\it primitive\/} generator, the unique indivisible element of $N$ which
spans the edge.
Then to each $v\in\S(1)$ is associated a divisor
$D_v\subset X_{\S}$.
Note that in general, $\S$ is not determined by $\S(1)$;
there may be several choices of fans with the
same set of edges.  In the applications we are interested in, this corresponds
to making a choice of phase of the same physical theory; or equivalently a
particular phase of the extended moduli space. One such phase will be the
usual region in which the  non-linear sigma model with a Calabi-Yau
manifold as the target space exists and another may be the corresponding
\LGO\ part.

As an example, we quickly review how to describe a product of weighted
projective
spaces as a toric variety using the fan construction.
For a single $r-1$ dimensional
weighted projective space $\cp{(k_1,\ldots,k_r)}{r-1}$, we put
$\vec{k}=(k_1,\ldots,k_r)$ and
take the $r-1$ dimensional
lattice
\eqn\eN{N=\ZZ^r/(\ZZ\cdot \vec{k}).}
Letting $e_1,\ldots,e_r$ be the standard basis for $\ZZ^r$, we build up
a fan from the edges $v_i$ spanned by the
images of the $e_i$ in $N$.
The cones in the fan are just the (simplicial)
cones spanned by all proper subsets of $v_1,\ldots,v_r$.
The toric variety associated to this fan
is $\cp{(k_1,\ldots,k_r)}{r-1}$.
The $U_\s$ corresponding to the $r$ top dimensional
cones are just the affine
open sets in $\cp{(k_1,\ldots,k_r)}{r-1}$ determined by the
condition that the coordinate corresponding to the edge not present in $\s$
is nonzero.

In the special case that $k_1=1$, then there is a natural
isomorphism $N\simeq\ZZ^{r-1}$ determined by projection onto the last
$r$ coordinates.  In these coordinates, the edges of the fan are spanned
by
\eqn\eWpfan{(-k_2,\ldots,-k_r),(1,0,\ldots,0),\ldots,(0,\ldots,1).}
For a product of weighted projective spaces (and more generally a product
of toric varieties), one forms a new lattice $N=N_1\times N_2$ by taking
a product of the lattices for the original varieties. (For simplicity
we consider here a product of two varieties; the construction goes through
for a product of any number of spaces.)
 The set of
all cones in $\S$
is simply the set of all products of a cone in the first fan $\S_1$ with
a cone in the second fan $\S_2$.  In particular, $\S(1)$
is the union $(\S_1(1)\times\{0\})\cup(\{0\}\times\S_2(1))$.
Said differently,
\eqn\eprodfan{\S(1)=\{\,\vec{v_i}\times\vec{0_2},\vec{0_1}\times\vec{w_j}\mid
              \vec{v_i}\in\S_1(1),\ \vec{w_j}\in\S_2(1)\,\}.}
Here, the $\vec{0_k}$ are the zero vectors in $N_k$ for $k=1,2$.

We turn next to the description of toric varieties via polyhedra.
Consider an $r$~dimensional integral polyhedron $P\subset M_{\IR}$,
whose vertices lie in
$M$.  One associates to $P$ an $r$~dimensional toric variety $\IP(P)$.
The construction also
gives a canonical embedding $\IP(P)\hookrightarrow\IP^{|P|-1}$, where
$|P|$ is the cardinality of $P\cap M$, and the coordinates of $\IP^{|P|-1}$
are identified with the points of $P\cap M$.

We note that $T$ may be embedded in $\IP^{|P|-1}$ via the map defined by
$t\mapsto (m_1(t),\ldots,m_{|P|}(t))$, where the $m_i$ range over the
points of $P\cap M$ (the $m_i$ are sometimes called monomials in this context).
The toric
variety $\IP(P)$ is in fact the closure of this map.  We will often think of
$\IP(P)$ as a projective variety in this fashion without further comment.

Conversely, consider a rational embedding of a weighted projective space
$\cp{(k_1,\ldots,k_r)}{r-1}$ into a projective space by the formula
$y_i=m_i(x)$, where $m_i(x)$ run over all monomials of fixed degree in
$\cp{(k_1,\ldots,k_r)}{r-1}$
(or more generally, of fixed multidegree in a product
of weighted projective spaces).  By restricting to the torus in our description
of $\cp{(k_1,\ldots,k_r)}{r-1}$ as a toric variety, the $m_i(x)$ restrict to
characters $m_i$ on the torus, which we identify as points of $M$.  The set
of all these lattice points span a polyhedron $P$, and we can recover a
blowup of the original
weighted projective space together with its embedding as the
toric variety $\IP(P)$.

\subsec{Batyrev's construction}
We next review Batyrev's proposed toric construction of mirror pairs
\rBatdual.\ft{In the case of  Fermat hypersurfaces in a weighted projective
space this toric construction was first noted by Roan~\rRoan.} Some aspects
of this construction have been amplified in~\rmdmm.

A {\it reflexive} polyhedron is an integral polyhedron $P$ containing 0 in its
interior, such that each facet of $P$ (that is, a codimension 1 face of $P$)
is supported by a hyperplane $H$ which can be defined by a linear equation
of the form $H=\{\,y\in M_{\IR}\mid \langle\ell,y\rangle =-1\,\}$ for some
$\ell$ in $N$.
Batyrev shows that if $P$ is reflexive, then
the general hyperplane section of $\IP(P)$ is \CY\ (possibly with mild
singularities, which can be resolved to obtain a \CY\ manifold).
The {\it polar polyhedron} (which in~\rBatdual\ is
called the dual polyhedron) is given by
\eqn\polar{P^\circ=\{\,x\in N_{\IR}\mid\langle x,y\rangle\ge -1\
\hbox{for all } y\in P\,\},}
and is reflexive if and only if $P$ is reflexive.  Batyrev proposes that the
hyperplane sections $\Mbar$ of $\IP(P)$ and $\Wbar$
of $\IP(P^\circ)$ should form a
mirror pair.

Furthermore, it turns out that $\IP(P^\circ)$ is also the toric variety
associated to the {\it normal fan\/} of $P$ (this is the fan whose cones are
simply the cones over the faces of $P$).  This observation leads immediately
to the monomial-divisor mirror map of~\rmdmm\ since certain points of
$M$ correspond simultaneously to edges of a fan, hence divisors on
(a partial desingularization of) $\IP(P^\circ)$ as well as monomials on
$\IP(P)$.

\noindent
{\bf Example:} Consider Calabi-Yau hypersurfaces (of degree~10) ${\cal M}$
in the weighted projective space $\cp{(1,2,2,2,3)}{4}$.  Note that by taking a
particular small radius limit of ${\cal M}$ we obtain the \LGO\ discussed
in section~2.3.
The above prescription
tells us to restrict to our affine piece by setting the first coordinate
equal to~1.  This identifies the 87~monomials of degree~10 with lattice
points by just looking at the last 4~exponents.  The extreme monomials are
seen to be
\eqn\eextmons{x_1^{10},\ x_1x_5^3,\ x_2^5,\ x_3^5,\ x_4^5,\ x_2^2x_5^2,\
              x_3^2x_5^2,\ x_4^2x_5^2.}
Looking at exponents and dropping the first coordinate, we see that $(1,1,1,1)$
(corresponding to the monomial $x_1x_2x_3x_4x_5$) is an interior point.
Translating this point to the origin, we see that $P$ is the convex hull of the
points
\eqn\everts{\matrix{(-1, -1, -1, -1) & (-1, -1, -1, 2) & (4, -1, -1, -1) \cr
          (-1, 4, -1, -1) & (-1, -1, 4, -1) & (1, -1, -1, 1) \cr
          (-1, 1, -1, 1) & (-1, -1, 1, 1)}}
The polar polytope $P^\circ$ is computed to be the convex hull of the
set of points
\eqn\epolarverts{\matrix{(-2, -2, -2, -3) & (1, 0, 0, 0) & (0, 1, 0, 0)\cr
          (0, 0, 1, 0) & (0, 0, 0, 1) & (-1, -1, -1, -2)}}
We see from~\eWpfan~that that $\cp{(1,2,2,2,3)}{4}$
is a toric variety associated to a fan with edges
\eqn\epfan{(-2,-2,-2,-3),\ (1,0,0,0),\ (0,1,0,0),\ (0,0,1,0),\ (0,0,0,1).}
Since $\IP(P)$ is determined by the normal fan associated to $P^\circ$, we
note as a check that $\IP(P)$ is birational to $\cp{(1,2,2,2,3)}{4}$
(the insertion of the additional edge $(-1,-1,-1,-2)$ effects a blowup).
Alternatively, we can consider $\IP(P^\circ)$ as the toric variety
associated to a certain fan with edges spanned by \everts.

We will return to this example later after discussing the toric explanation
of the transposition rule.

\subsec{The construction of Batyrev and Borisov}

There is a general proposed construction of mirror pairs for complete
intersections in certain toric varieties by Borisov~\rBor.  This can
also be explained in terms of a later construction of Batyrev and
Borisov~\rBatBor\ which allows one to also see the Landau-Ginzburg phase
of the same theory in case such a phase exists.
For this reason, we primarily focus on
explaining the  approach taken in~\rBatBor.  The analysis
of phases is essentially the same as described in~\refs{\rphases,\rAGM}.

We consider again a reflexive polyhedron $P\subset M_{\IR}$ and its normal
fan $\S\subset N_{\IR}$.  Recall that the edges $v_j\in\S(1)$
correspond to
divisors $D_{v_j}$ in $\IP(P^\circ)$, and that the hyperplane sections are in
the same divisor class as $\sum_jD_{v_j}$.

For complete intersections, we consider a {\it nef partition\/} of the edges
of $\S$.  That is, we decompose $\S(1)$ into a disjoint union
$\cup_{i=1}^k\S(1)_i$ of nonempty subsets of $\S(1)$.  For each $i$,
the divisor class $H_i=\sum_{v_j\in\S(1)_i}D_{v_j}$ is assumed to be nef
(semi-ample), which means that $H_i\cdot C\ge 0$ for all curves $C$.
To this data, one can associate mirror families of Calabi-Yau complete
intersections of codimension $k$.  The $k$ hypersurfaces comprising the
complete intersection are to lie in the respective divisor classes $H_i$.

Put $\bN=N\oplus\ZZ^k$ and let $\bM$ be the dual lattice,
with $\bN_{\IR}$ ($\bM_{\IR}$) defined in the obvious way.
Put $\bar{v_j}=(v_j,e_i)\in\bN$ whenever $v_j\in\S(1)_i$.
Let
\eqn\erefcone{\s=\IR_{\ge 0}\langle \bar{v_1},\ldots,\bar{v_k},
                 (0,e_1),\ldots,(0,e_k)\rangle\subset\bNR.}
Here the $e_i$ are the standard basis elements of $\ZZ^k$.

Consider the dual cone
\eqn\edualcone{\check\s=\{\,y\in\bMR\mid \langle x,y\rangle \ge 0\ {\forall}\
x\in\s\}.}
One can again find a nef partition in $M$ for which application of the
preceding construction
yields $\check\s$.  One supposes that the two
families of complete intersections so obtained are mirror families.

As in the previous subsection, one can associate monomials to certain points of
$\s\cap\bN$ and $\check\s\cap\bM$ respectively.

\noindent
{\bf Example}. Let us consider $\IP^5[2,4]$ which
was first studied in the context of mirror symmetry  in~\rLib, (see also \rKT).
By~\eWpfan,
$\IP^5$ may be realized as the toric variety associated to the fan with
edges spanned by the vectors $(-1,-1,-1,-1,-1),\ (1,0,0,0,0)$, $(0,1,0,0,0),\
(0,0,1,0,0),\ (0,0,0,1,0)$, $(0,0,0,0,1)$.  We take $x_0,\ldots,x_5$ as
homogeneous coordinates on $\IP^5$.  The edges given above correspond to
the hyperplanes $x_0=0,\ldots,x_5=0$.

We choose a nef partition by dividing the set of edges into two
subsets---the first two and the last four.  This corresponds to the choice
of quadric and quartic  hypersurfaces comprising the
complete intersection.

The cone $\s$ is spanned by the vectors
\eqn\epcone{\matrix{(-1,-1,-1,-1,-1,1,0) & (1,0,0,0,0,1,0) & (0,1,0,0,0,0,1)\cr
          (0,0,1,0,0,0,1)      & (0,0,0,1,0,0,1) & (0,0,0,0,1,0,1)\cr
          (0,0,0,0,0,1,0)      & (0,0,0,0,0,0,1) }}
The dual cone $\check\s$ is similarly associated to the mirror
manifold.  The edges of $\check\s$ are computed to be
\eqn\edualpcone{\matrix{(-1,0,0,0,0,1,0) & (1,0,0,0,0,1,0) &
                                                     (-1,2,0,0,0,1,0)\cr
          (-1,0,2,0,0,1,0) & (-1,0,0,2,0,1,0) &(-1,0,0,0,2,1,0)\cr
          (0,-1,-1,-1,-1,0,1) & (4,-1,-1,-1,-1,0,1) & (0,3,-1,-1,-1,0,1) \cr
          (0,-1,3,-1,-1,0,1)  & (0,-1,-1,3,-1,0,1)  & (0,-1,-1,-1,3,0,1) \cr
          (0,0,0,0,0,1,0) & (0,0,0,0,0,0,1)}}
Later, in section~3.5, we will show that the mirror family
appears as a complete intersection in an orbifold of $\IP^5$, as asserted
in~\rLib.  For now, we note how to see the monomials
comprising the original complete intersection.  The last two coordinates
distinguish the first 6~edges from the next~6 (the last two play a separate
role, and may be ignored for our current purposes).
The first 6~edges form a 5~dimensional
simplex of side~2
corresponding to the quadrics; the next 6~edges form a 5~dimensional
simplex of side~4
corresponding to the quartics.

The exact correspondence can be described as follows.
The extreme vertices of the set of polynomials of
degree~2 are the $x_i^2$; those for degree~4 are similarly the $x_i^4$.
They are naturally identified as polynomials of degree~6 (the anticanonical
series of $\IP^5$) by multiplying all quartics by $x_0x_1$ (this coming
from the partition) and multiplying all quadrics by $x_2x_3x_4x_5$.
This gives the following list of monomials of degree~6.
\eqn\ecimons{\matrix{x_0^2x_2x_3x_4x_5 & x_1^2x_2x_3x_4x_5 & x_2^3x_3x_4x_5\cr
          x_2x_3^3x_4x_5    & x_2x_3x_4^3x_5 & x_2x_3x_4x_5^3\cr
          x_0^5x_1          & x_0x_1^5       & x_0x_1x_2^4\cr
          x_0x_1x_3^4       & x_0x_1x_4^4    & x_0x_1x_5^4 }}
The identification with the
vectors in~\edualpcone\ comes from
ignoring the $x_0$ factors
in~\ecimons, then
shifting all remaining exponents by $-1$.  This gives the first 5~coordinates
of~\edualpcone.
In this way, the monomials occur
naturally inside the polar polytope $P^\circ$ as in the hypersurface
case~\rBor.  Alternatively, one can keep all coordinates in~\edualpcone\ and
use the last two coordinates to distinguish which factor of the complete
intersection the monomial lies in~\rBatBor.  In this interpretation, we need
not multiply by $x_0x_1$ or by $x_2x_3x_4x_5$.

It remains to relate this to the phases analysis of~\rphases.  Standard
procedures for rewriting toric varieties as torus quotients show that
we must mod out $\IC^8$ by the $\IC^*$ action with weights
$(1,1,1,1,1,1,-2,-4)$ (these computed by noting that the eight edges found
in~\epcone\ sum to zero after being given the above coefficients).
Translating into the language of~\rphases,
we consider the superfields $x_0,\ldots,x_5,p_1,p_2$ where the
$x_i$ have weight~1, $p_1$ has weight $-2$ and $p_2$ has weight $-4$.
Form the superpotential $p_1f(x)+p_2g(x)$ which has weight~0.  Witten's
phases analysis coincides with the decomposition via the secondary fan
of the points of~\epcone;
either way, there are two phases: the Calabi-Yau phase and
a hybrid of a \CY\ and a \LGO\ phase, with a gauged \LG\ theory as
the boundary between the two.
In  a more general situation there will be more than
the (gauged) Landau-Ginzburg and the Calabi-Yau theories as well
as various hybrids of \CY\ and \LGO\ phases. In fact in the above example
one can show that there does not exist a Landau-Ginzburg phase.

\subsec{Transposition via toric geometry}\noindent
In this subsection we sketch the toric explanation of the transposition rule
of~\rBH; for more details, see~\rCOK.  For convenience, we will study
4~dimensional weighted projective spaces, but that restriction is not
essential.

Consider the weighted projective space $\wp$,
and let $d=\sum_i k_i$.  As we have seen from our earlier discussion,
we can describe the family of Calabi-Yau hypersurfaces in terms of the
Newton polyhedron $P\subset M$ spanned by all monomials of degree $d$,
translating the monomial $x_1\cdots x_5$ to the origin.

Suppose further that from among the degree $d$ monomials in
$\cp{(k_1,k_2,k_3,k_4,k_5)}{4}$ one is given
5~monomials $m_1,\ldots,m_5$. These monomials
will sometimes be identified with the
corresponding lattice point in $P$.  We assume that $P$ is reflexive.
Let us also assume
that the $m_i$ span $M$.
Note that we do not require that the
general polynomial formed from these 5~monomials be transverse, which
would have been the case had we tried to formulate the theory in terms
of a Landau-Ginzburg orbifold.

 From these monomials, we form  the matrix $A$ of
exponents of the terms of the polynomial $p=\sum_im_i$ (see e.g.~\epmat).
Recall that $A$ is a $5\times 5$ matrix  with
the exponents of each monomial  in the columns of $A$.

Our assumptions imply that
there exists a relation between the $m_i$ such that
\eqn\enewwts{
\sum_{i=1}^5\khat_im_i=\vec{0}
}
for some integers $\khat_i$ which are well-defined up to an overall
scale.
Finally, we assume that the $m_i$ do not all lie on
the same side of any hyperplane passing through the origin. But this
implies that the $\khat_i$ all have the same sign,
and in particular may be chosen to all be
positive.

With these assumptions,
our assertion is that the mirror manifold is obtained from
the original equation by the transposition rule.  That is, one transposes
$A$ to get $5$ new monomials in
$\cp{(\khat_1,\khat_2,\khat_3,\khat_4,\khat_5)}{4}$,
forms their
sum to get the transposed polynomial $\phat$, takes an appropriate
orbifold, and resolves singularities to get the mirror manifold.
So far this is nothing new compared to the original construction
of~\rBH, since the monomials arising from~\etadloop{} may be seen to give the
required properties; in particular $P$ is reflexive.\ft{In certain cases
 an ambiguity occurs in resolving the
singularities; the model as it is defined in terms of the relevant
quotient of the transposed polynomial is at a boundary point at which
two components of the moduli space corresponding to two distinct \CY\
manifolds meet.
However, only one of them is the mirror to the original model~\rCOK.}

We now recall the lattice points $v_i\in N$ induced by the standard basis
vectors $e_i$.
It turns out that $P^\circ$ contains the six
points ${\vec 0},v_1,v_2,v_3,v_4,v_5$~\rBK.
To  ${\vec 0},v_i$ correspond
monomials $\hat m_i$
 in the toric variety determined by $P^\circ$, and one obtains
a polynomial $\hat{p}=\sum_i\hat{m_i}$.  Since the
$v_i$ are linearly dependent, with the only relation being
$\sum_ik_iv_i=0$, we have
\eqn\erelate{
\prod_{i=1}^5 \hat m_i^{k_i}=\hat m_0^d~.}
Here $\hat m_0$ corresponds to ${\vec 0}$ and
it is easy to see that all relations between the $\hat m_i$ are just powers
of~\erelate.

By comparing the lattice in the toric description of
$\cp{(\khat_1,\khat_2,\khat_3,\khat_4,\khat_5)}{4}$
(see~\eN\ with $\vec{k'}$ in place of $\vec{k}$)
with~\enewwts, we see that there is a map of fans induced from the
mapping of lattices
\eqn\emapfan{\ZZ^5/(\ZZ\cdot\vec{k'})\to M}
which takes $v_i$ to $m_i$ (the normal fan of $P^\circ$ may need to be
subdivided so that the $m_i$ span a cone; this amounts to a choice of phase).
This gives $\IP(P^\circ)$ birationally as an orbifold of
$\cp{(\khat_1,\khat_2,\khat_3,\khat_4,\khat_5)}{4}$.

We can now observe that when
referred back to $\cp{(\khat_1,\khat_2,\khat_3,\khat_4,\khat_5)}{4}$
via the orbifold determined
by~\emapfan,
$\hat p$ is just the transposed polynomial, as has been asserted earlier.
To see this, it suffices to observe that the entries of the matrix $A$
are obtained by adding~1 to the inner products of the $m_i\in M$ with the
$v_j\in N$.  Mirror symmetry switches the roles of the $m$'s and the
$v$'s, resulting in the transposition of the matrix.  This will be amplified
in the examples below.

The final step  is to verify that the group of geometric and
quantum symmetries for the model and its mirror have
the claimed order.  Of course, the toric method gives the group explicitly.
\ft{Although not stated in~\rBH\
it was recently shown that the explicit quotient on
the transposed model can be obtained from the usual transposition
analysis~\refs{\rPM,\rMax}.}
The calculation is carried out by the standard toric
description of the order of finite quotient mapping in terms of the index
of a certain sublattice.  An example will be
given presently.

\subsec{Examples}

\noindent
Let us now return to the example of hypersurfaces of degree ten
in $\cp{(1,2,2,2,3)}{4}$.
Consider the monomials
\eqn\esomemons{x_1^{10},\ x_1x_5^3,\ x_2^5,\ x_3^5,\ x_4^5.}
Taking these for our $m_i$ and using the earlier conventions about
coordinates for our lattices, we identify these with
the first~5 points from~\everts,
which we will reorder slightly:
\eqn\emons{\matrix{(-1, -1, -1, -1) & (4, -1, -1, -1) & (-1, 4, -1, -1) \cr
                  (-1, -1, 4, -1) & (-1, -1, -1, 2) }}
These points satisfy $m_1+3m_2+3m_3+3m_4+5m_5=0$. Hence, from~\enewwts\
and the subsequent discussion, the mirror is
an orbifold of $\cp{(1,3,3,3,5)}{4}[15]$ where we let $(y_1,\ldots,y_5)$ denote
coordinates on $\cp{(1,3,3,3,5)}{4}[15]$.

To see this more clearly, the 5~points above may be thought of as points of
$P\subset M$,
or as edges of the normal fan of $P^\circ$.  In the latter interpretation,
these edges span a cone in the subdivided fan (recall that the fan gets
subdivided to partially resolve singularities), and we can map
$\cp{(1,3,3,3,5)}{4}$ to it by the mapping of fans induced by the natural
linear mapping
\eqn\efanmap{\ZZ^5/(\ZZ\cdot(1,3,3,3,5))\to M}
which sends the $e_i$ to $m_i$, as in~\emapfan.
Now the $v_i\in N$ may be interpreted as
monomials on $\IP(P^\circ)$; the exponent of the variable $y_j$ in this
monomial is obtained by adding~1 to $\langle v_i,m_j \rangle$
because of the way that the
mapping from $\cp{(1,3,3,3,5)}{4}$ has been defined
(recall that
in our coordinates, $v_1=(-2,-2,-2,-3)$).  On the other hand,
$\langle v_i,m_j \rangle +1$ is also the exponent of $x_i$ in the monomial
$m_j$.  Thus, the exponents of the $y_j$
are clearly given by the transposition rule.

Finally, we have to check that~\eQS{} are fulfilled.
To do so consider a one parameter family of deformations
${\cal W}=\tilde {\cal W}/G$ where $\tilde {\cal W}
 \in \cp{(1,3,3,3,5)}{4}[15]$ is defined by
\eqn\edefex{
\hat p_\j=\hat p_0 - 10 \j  \prod_{i=1}^5 y_i~
=y_1^{10}y_5+y_5^3+y_2^5+y_3^5+y_4^5 - 10 \j \prod_{i=1}^5 y_i,}
and $G=(\ZZ_5:4,1,0,0,0)\times (\ZZ_5:4,0,1,0,0)$.
Here $\hat p_0$ is the transpose of
$p_0=x_1^{10}+x_1x_5^3 + x_2^5 + x_3^5 + x_4^5$, where $p_0=0$
is the mirror ${\cM}\in\cp{(1,2,2,2,3)}{4}[10]$ of ${\cal W}$ (see also the
discussion at the end of section~2.3).
Following~\rCdGP, we can desingularize this orbifold by identifying
singularities.  Below is a list of equations of curves in the hypersurface
together with an automorphism that fixes the curve.
\eqn\efixcurve{\matrix{x_1=x_2=0 & (4,1,0,0,0) \cr
                       x_1=x_3=0 & (4,0,1,0,0) \cr
                       x_1=x_4=0 & (4,0,0,1,0)\cr
                       x_1=x_5=0 & (\ZZ_3 : 1,0,0,0,2) \cr
                       x_2=x_3=0 & (0,1,4,0,0) \cr
                       x_2=x_4=0 & (0,1,0,4,0) \cr
                       x_3=x_4=0 & (0,0,1,4,0) \cr}}
With one exception, the
automorphism comes from $G$.  For $x_1=x_5=0$, the automorphism comes from
the definition of the weighted projective space.
These curves intersect in the following collection of finite point sets.
\eqn\efixpts{\matrix{x_1=x_2=x_3=0 & x_1=x_2=x_4=0 & x_1=x_2=x_5=0 \cr
                     x_1=x_3=x_4=0 & x_1=x_3=x_5=0 & x_1=x_4=x_5=0 \cr
                                   & x_2=x_3=x_4=0 &               \cr}}
In addition, the point $(1,0,0,0,0)$ is singular, since it lies on the
singular set $x_2=x_3=x_4=0$ and also on {\it every\/} member of our family
$\hat p_\j$.

We calculate the Euler characteristic of the Calabi-Yau resolution of
${\cal W}$
as usual by finding the isotropy groups of the
points and curves.
{}From this, the Euler characteristic of the Calabi-Yau resolution
can be computed to be 144, which is indeed the negative of the Euler
characteristic of ${\cal M}$.

We want to relate~\edefex\
to the corresponding realization of ${\cal W}$ as a hypersurface in a toric
variety given by
\eqn\etorfam{
\sum_{i=1}^5m_i-\psi dm_0=0~.}
To do so  we make the identifications
\eqn\eident{
\hat m_0=y_1y_2y_3y_4y_5\,,\quad \hat m_1=y_1^{10}y_5\,,
\quad \hat m_i=y_i^5\,,i=2,3,4
\,,\quad \hat m_5=y_5^3~}
as required by the preceding discussion;
note that $\hat m_1\hat m_2^2\hat m_3^2\hat m_4^2\hat m_5^3=\hat m_0^{10}$
 from~\erelate. This map is not well defined since
 the $\hat m_i$ are invariant under
a $\ZZ_3\times\ZZ_5^3$ action generated by
$G\times(\ZZ_{15}:1,3,3,3,5)$ on the $y_i$; $\hat m_0$ transforms under a
$\ZZ_{10}$ under the group of rescalings of the $y_i$ which preserve
the $m_i$ for $i\neq 0$.
Thus, to make the identification $1-1$ we have to consider a quotient
by $G\times\ZZ_{15}$. But the $\ZZ_{15}$ is already enforced by the
projectivization in $\cp{(1,3,3,3,5)}{4}[15]$. Hence, we are left with
$\cp{(1,3,3,3,5)}{4}[15]/G$ which is the result obtained using
the transposition argument.

We can also resolve singularities of ${\cal W}$ by toric geometry,
subdividing the normal fan of $P^\circ$ by including new edges for all nonzero
lattice points of $P$.  By doing this, we in fact get more loci to blow up
than given by~\efixcurve, \efixpts, and $(1,0,0,0,0)$.  The toric method
suggests that the points $(0,1,0,0,0),\ (0,0,1,0,0),\ (0,0,0,1,0),\
(0,0,0,0,1)$ should be resolved.  The reconciliation of the two procedures
is that the general member of $\hat p_\j$ does not contain any of these
4~points, so resolving this point in the toric 4~fold does not affect
the Calabi-Yau threefold.

This example illustrates that the transposition scheme can often be used
independent of toric methods, yielding the same result.
Toric methods have the advantage of being more general and are easier
to use for those familiar with toric techniques.

The extension to models for which one (or both) of $p$ and $\hat p$ are not
transverse polynomials is straightforward along the lines described above.
However, we cannot compare with the original construction since the
Landau-Ginzburg formalism does not yet extend to the more  general
picture for which toric considerations apply (see also~\rCOK,\rBatdual).
However, we can compare the construction to more general
Landau-Ginzburg theories with more than 5~superfields as well as to complete
intersection Calabi-Yau manifolds when a Landau-Ginzburg description
exists. We will next turn to an example of the latter kind.

Consider the example
\eqn\eci{\K{4\cr1\cr}{4&1\cr0&2\cr}^{2,86}_{-168}~~:~~
 \cases{f(x_a)   = f_{abcd}\, x_a\, x_b\, x_c\, x_d &= 0~,\cr
        g(x_a,x_\a) = g_{a\,\a\b}\, x_a\, x_\a\, x_\b  &= 0~,\cr}}
which is a complete intersection of a quartic polynomial, $f(x_a)$
in the $\CP{4}$
variables $x_1,\ldots,x_5$
and another polynomial, $g(x_a,x_\a)$ which is linear in the $\CP{4}$ variables
and quadratic in the $\CP{1}$ variables $x_6,x_7$.  An example is given by
the polynomials
\eqn\equartic{x_1^4+x_2^4+x_3^4+x_4^4+x_5^4}
and
\eqn\eonetwo{x_1x_6^2+x_2x_7^2.}
We may add these potentials to get a new potential
\eqn\eLG{p(x)=x_1^4+x_2^4+x_3^4+x_4^4+x_5^4+x_1x_6^2+x_2x_7^2,}
which may be interpreted as a Landau-Ginzburg potential for the Landau-Ginzburg
orbifold by $(\ZZ_8:1,1,1,1,1,4,4)$.
This theory is in the same moduli space as the Calabi-Yau theory described
by the complete intersection~\eci, but in a different
phase~\refs{\rphases,\rAGM}.

We will now construct the mirror model of ${\cal M}$ thought of either
as a complete intersection or as a Landau-Ginzburg orbifold and show
that these mirrors are different phases of the same theory. Note that for
the generic complete intersection \CY\ it is in general
not possible to associate
a \LG\ potential as we have done in~\eLG, \ie\ there does not exist a
\LG\ phase and hence  in the mirror moduli space there will not be
a \LG\ phase either~\rphases.

Using the transposition technique described in section~3, we obtain
the transposed polynomial as (recall that transposition applies for
any number of fields and any central charge modulo certain conditions on
the non-degeneracy of the superpotential~\rCOK)
\eqn\LGtr{p^T(y)=y_1^4 y_6 + y_2^4 y_7 + y_3^4 + y_4^4 + y_5^4 +
y_6^2 + y_7^2.}
The corresponding Landau-Ginzburg orbifold is given in terms of a
$(\ZZ_8:1,1,2,2,2,4,4)$ quotient of $p^T$. In order for~\eQS{} to hold
we need to further divide by $H=(\ZZ_4)^3$; note that $Q=\ZZ_8$ and
$G=\ZZ_8\times(\ZZ_4)^3$ for both $p/\ZZ_8$
and $p^T/\ZZ_8$.

We now apply the construction of Batyrev and Borisov~\rBatBor.
The fan for $\IP^4\times\IP^1$ has edges
\eqn\efouroneed{\matrix{(-1,-1,-1,-1,0)&(1,0,0,0,0)&(0,1,0,0,0)&(0,0,1,0,0)\cr
          (0,0,0,1,0)&(0,0,0,0,-1)&(0,0,0,0,1)\cr}}
To form the complete intersection, these edges are partitioned into
two sets:
\eqn\eedgea{\{(1,0,0,0,0),(0,1,0,0,0),(0,0,1,0,0),(0,0,0,1,0)\}}
and
\eqn\eedgeb{\{(-1,-1,-1,-1,0),(0,0,0,0,-1),(0,0,0,0,1)\}}
corresponding to $f(x_a)$ and $g(x_a,x_\a)$ respectively in~\eci.  The
cone is formed by appending $1,0$ to the edges~\eedgea, and
appending $0,1$ to the edges~\eedgeb.  This gives
\eqn\efouronecone{\matrix{(1,0,0,0,0,1,0)&(0,1,0,0,0,1,0)&(0,0,1,0,0,1,0)\cr
                        (0,0,0,1,0,1,0)&(-1,-1,-1,-1,0,0,1)&(0,0,0,0,-1,0,1)\cr
                                       &(0,0,0,0,1,0,1)\cr}}
As before, we can omit the two vertices $(0,0,0,0,0,1,0)$ and
$(0,0,0,0,0,0,1)$ for our immediate purposes.
These edges determine an affine seven-dimensional toric variety.  The
magnitude of the determinant of the $7\times 7$ matrix formed from these
vectors is
8; this is identified with the $(\ZZ_8: (1,1,1,1,1,4,4))$ of the associated
Landau-Ginzburg orbifold.

We now calculate that the dual cone is spanned by 15~edges.  These edges are
spanned by the following 15~vectors (omitting $(0,0,0,0,0,1,0)$ and
$(0,0,0,0,0,0,1)$):
\eqn\edualedge{\matrix{(-1,-1,-1,-1,0,1,0) & (3,-1,-1,-1,0,1,0)
                                                 & (-1,3,-1,-1,0,1,0)\cr
           (-1,-1,3,-1,0,1,0) & (-1,-1,-1,3,0,1,0) & (0,0,0,0,-1,0,1)\cr
           (0,0,0,0,1,0,1)    & (1,0,0,0,-1,0,1)   & (1,0,0,0,1,0,1)\cr
           (0,1,0,0,-1,0,1)   & (0,1,0,0,1,0,1)    & (0,0,1,0,-1,0,1)\cr
           (0,0,1,0,1,0,1)    & (0,0,0,1,-1,0,1)   & (0,0,0,1,1,0,1)}}
According to~\rBatBor, these edges have two interpretations:
first as edges of a cone describing the mirror, and second, as data
corresponding to monomials determining the complete intersection in the
original variety.  Let us check the latter interpretation.
The last two coordinates partition this set into two sets: the first
five and last ten.  The projection of the
first five vectors onto the first five coordinates span a 4~dimensional
simplex of side~4 (most easily
seen by translating the first vertex to the origin); thus the convex hull
of these may be easily identified with the quartic monomials in $\IP^4$.
The projection of the last 10 onto the first five coordinates
constitute a product
\eqn\eprod{\{(0,0,0,0),(1,0,0,0),(0,1,0,0),(0,0,1,0),(0,0,0,1)\}\times
            \{(-1),(1)\}}
This is a product of a 4-dimensional simplex of side~1 with a one-dimensional
simplex of side~2.  The convex hull of this set is thus naturally identified
with the monomials which are linear in $\IP^4$ and quadratic in $\IP^1$.

More interestingly, let us look at the first interpretation.  We can
realize the toric variety associated to our cone as a quotient of $\IC^7$
 in a particularly nice way.  We pick seven of our edges
\eqn\eLGedges{\matrix{(3,-1,-1,-1,0,1,0) & (-1,3,-1,-1,0,1,0)
                                                & (-1,-1,3,-1,0,1,0) \cr
          (-1,-1,-1,3,0,1,0) & (-1,-1,-1,-1,0,1,0)& (1,0,0,0,-1,0,1)   \cr
                             & (0,1,0,0,1,0,1)}}
These edges have been chosen to correspond to~\eLG\ under our identifications,
\ie\ with $x_i$ being the coordinates of the toric orbifold
represented by~\efouronecone\ the exponent of $x_i$ in a monomial is
obtained by taking the inner product of the vector from~\efouronecone\ with
the vector from~\eLGedges. This gives us the monomials in $p$,~\eLG.
The vectors~\eLGedges{} are linearly independent, and the $7\times 7$
 matrix that
they form has
determinant~512.  Thus there is a rational map from $\IC^7$ to our toric
variety which is generically 512~to~1 (the cone must be subdivided to
allow the span of these seven edges to be a cone in the subdivided fan;
this corresponds to a choice of phase in the theory).  Before we interpret
this quotient, let us look at the monomials, which correspond to the original
7~edges~\efouroneed.  Let $y_1,\ldots,y_7$ be
coordinates on $\IC^7$.  To describe the monomials in these coordinates, we
repeat the above procedure with the difference that
 the exponent of $y_i$ in a monomial of the transposed polynomial
is  the inner product
of the  vector from~\eLGedges\ corresponding to $y_i$ with the
vector from~\efouronecone\ corresponding to
the monomial on the mirror.  The result is the list of monomials
\eqn\emonlist{y_1^4y_6,\ y_2^4y_7,\ y_3^4,\ y_4^4\ y_5^4,\ y_6^2,\ y_7^2,}
agreeing with the result obtained by transposition.  Note also that these
monomials respect the group of order 512 generated by
$\ZZ_8\times\ZZ_4^3$, so must coincide with the quotient taken here.

Finally, let us return to the \CY\ hypersurfaces given by
$\IP^5[2,4]$.
 From the list~\edualpcone\ of edges of the dual cone $\check\s$
calculated earlier, we
choose the following 6~vectors:
\eqn\eorbedge{\matrix{(0,-1,-1,-1,-1,0,1) & (4,-1,-1,-1,-1,0,1)
                                          & (-1,2,0,0,0,1,0) \cr
         (-1,0,2,0,0,1,0) & (-1,0,0,2,0,1,0) & (-1,0,0,0,2,1,0)}}
The sum of the first 5~coordinates of
these vectors is zero; hence there is a rational mapping from
$\IP^5$ to the toric variety in which the mirror family is contained
(recall from~\rBatBor\ that the fan of this toric variety contains
{\it all\/} of the
(first 5~coordinates of) the edges).
 From this data, the description of the mirror family as a complete
intersection
in an orbifold of $\IP^5$ can be recovered.  In fact, by taking inner products
of~\eorbedge\ with~\epcone\ we get the monomials
\eqn\eLTmons{y_1^4,\ y_2^4,\ y_3^4,\ y_4^4,\ y_5^2,\ y_6^2,\ y_3y_4y_5y_6,
              \ y_1y_2}
which correspond to those used in~\rLib.

\newsec{Discussions}\noindent
Despite a fair amount of success in the construction of potential mirror
pairs described in this article,
\ie\ models ${\cal M}$ and ${\cal W}$ for which the spectrum is
flipped and~\eQS{} is fulfilled, the main question is still left: how
do we prove the construction at the level of the conformal field
theory?  Recently, progress has been made in understanding the moduli
space in terms of toric varieties; for the parts of the moduli space
which can be described as toric varieties it was shown that the
enlarged K\"ahler moduli space of the original manifold is
isomorphic to the complex structure moduli space of its
mirror~\rAGM. (For more details, see other
articles in this volume.) Although there is compelling evidence for the mirror
symmetry constructions as reviewed above, it does not account for
those parts of the moduli space which are not described by means of toric
geometry. Also, it is still only a
construction at the level of algebraic geometry and does not tie
together the conformal field theories of ${\cal M}$ and ${\cal W}$.

Another approach is to try to understand the $N=2$ Landau-Ginzburg orbifolds
better and that way get insight into the conformal field theory to which
these theories flow in the IR-limit. In particular properties which
are independent of the renormalization group flow, such as the elliptic
genus, have been computed. For the class of models considered in~\rBH\
the calculation of the elliptic genus~\rPM\ as well as the
study of the ring structures~\rMax\ supports the mirror symmetry
conjecture between a model and a particular orbifold of its transpose.
It would be interesting to extend this work to the remaining
models for which the Landau-Ginzburg approach fails.

In the context of the Landau-Ginzburg models it is worth pointing out
a difference between the transposition construction as applied to these
models and the toric technique. While there has to exist a geometric
interpretation of the conformal field theory of choice for the toric
geometry to apply, \ie\ $\hat c\in\ZZ$, this is not the case for
the $N=2$ field theories; the transposition argument applies equally well
to any model. In fact this is the case already for mirror symmetry
applied to the $N=2$ minimal models. Thus, one may hope that
the toric methods could be applied to the Landau-Ginzburg type models as
well.

Finally, recall that, as defined in~\eACTION, the $N=2$ Landau-Ginzburg models
are only supersymmetric and not conformally invariant. Still, all of
the features we have relied on in classifying potential mirror pairs
only rely on the properties, like the discrete symmetries, the spectrum
and the elliptic genus, of the massive theory and not on the existence
of a conformal fixed point. This would indicate that mirror symmetry is
indeed a two-dimensional feature independent of any possible spacetime
interpretations.

\vskip 5mm

{\bf Acknowledgments}:
P.B. wishes to thank T.~H\"ubsch and M.~Henningson for fruitful collaborations
leading to results presented in this article.
P.B. was supported by  DOE grant DE-FG02-90ER40542.  S.K. was supported
by NSF grant DMS-9311386.

\vfill
\eject

\listrefs

\bye